\documentclass{aastex62}

\def\ga{\mathrel{\hbox{\rlap{\hbox{\lower4pt\hbox{$\sim$}}}\hbox{$>$}}}}
\def\la{\mathrel{\hbox{\rlap{\hbox{\lower4pt\hbox{$\sim$}}}\hbox{$<$}}}}

\usepackage{verbatim}
\usepackage{amsmath}
\usepackage{multirow}
\usepackage{relsize}
\usepackage{natbib}
\usepackage[flushleft]{threeparttable}
\usepackage{hyperref}
\usepackage{multirow}
\usepackage{array}

\usepackage{amssymb}
\usepackage{float}

\shorttitle{RadFil Python Package}

\begin{document}

\title{\texttt{RadFil}: a Python Package for Building and Fitting Radial Profiles for Interstellar Filaments}

\author{Catherine Zucker}
\affiliation{Harvard-Smithsonian Center for Astrophysics, 60 Garden St. Cambridge, MA 02138, USA}

\author{Hope How-Huan Chen}
\affiliation{Harvard-Smithsonian Center for Astrophysics, 60 Garden St. Cambridge, MA 02138, USA}

\collaboration{(co-PIs)}

\correspondingauthor{Catherine Zucker}
\email{catherine.zucker@cfa.harvard.edu}

\begin{abstract}
We present \href{https://github.com/catherinezucker/radfil}{\texttt{RadFil}}, a publicly available Python package that gives users full control over how to build and fit radial density profiles for interstellar filaments. \texttt{RadFil} builds filament profiles by taking radial cuts across the spine of a filament, thereby preserving the radial structure of the filament across its entire length. Pre-existing spines can be inputted directly into \texttt{RadFil}, or can be computed using the \href{https://github.com/e-koch/FilFinder}{\texttt{FilFinder}} package as part of the \texttt{RadFil} workflow. We provide Gaussian and Plummer built-in fitting functions, in addition to a background subtraction estimator, which can be fit to the entire ensemble of radial cuts or an average radial profile for the filament. Users can tweak parameters like the radial cut sampling interval, the background subtraction estimation radii, and the Gaussian/Plummer fitting radii. As a result, \texttt{RadFil} can provide treatment of how the resulting filament properties rely on systematics in the building and fitting process. We walk through the typical \texttt{RadFil} workflow and compare our results to those from an independent radial density profile code obtained using the same data; we find that our results are entirely consistent. \texttt{RadFil} is open source and available on \href{https://github.com/catherinezucker/radfil}{GitHub}. We also provide a complete working tutorial of the code available as a \href{https://github.com/catherinezucker/radfil/blob/master/RadFil_Tutorial.ipynb}{Jupyter notebook} which users can download and run themselves. 

\end{abstract}

\section{Introduction}
\label{intro}
The appearance of increasingly high-resolution continuum and spectral line observations (e.g. from Herschel, Spitzer, ALMA, APEX) over the last decade has revealed a wealth of filamentary structure in the interstellar medium of our Galaxy. For instance, Herschel observations of the dense material in nearby molecular clouds show bundles of filamentary complexes, the densest of which gravitationally fragment and undergo star formation \citep{Andre_2010, Schisano_2014}. On larger scales, imaging of the Galactic plane in both emission and absorption reveal 50-100 pc long grand design filaments, some of which appear to be associated with the Galaxy's spiral arms \citep{Goodman_2014,Zucker_2015, Wang_2015,Zucker_2017}. 

The ubiquity of filaments across the Galaxy has given rise to a number of filament finding algorithms \citep{Sousbie_2013,Koch_2015,Schisano_2014,Menshchikov_2013}, intended to detect and characterize the properties of filaments in an automated fashion. Two properties of prime interest are the widths and radial profiles of filaments, which have important implications for the dominant physics (gravity, turbulence, magnetic field orientation) involved in filament formation \citep[e.g.][]{Arzoumanian_2011,Seifried_2015}. Unfortunately, no publicly available code currently exists to investigate radial profiles of filaments in a systematic way. Most currently available algorithms adopt a broad-brush approach to building filament profiles, by assessing how the average intensity of the filament drops as a function of distance from the density ridge or ``spine" of the filament. However, this method fails to capture how the profiles of the filaments can change as a function of position along the spine, or be affected by density peaks (e.g. clumps, cores) offset from the main spine. 

To fill this void, we have created \texttt{RadFil}\footnote{The version of the \texttt{RadFil} code used to derive the results in this paper has been archived on Zenodo:\dataset[10.5281/zenodo.1287318]{https://doi.org/10.5281/zenodo.1287318}.}---a publicly available radial profile fitting code which provides a tailored approach to filament width analyses. Rather than estimating filament intensities within concentric distance bins from the spine, \texttt{RadFil} builds radial profiles by taking radial cuts across the spine, thereby preserving the morphology of the filament at every point along its length. We provide two built in fitting functions (Plummer and Gaussian), along with a background subtraction estimator, which can be applied to the average cut or ensemble of cuts across the filament. And because all the building and fitting knobs are tunable, \texttt{RadFil} can be applied to filaments of all size scales and densities, obtained via observations or numerical simualations. In \S \ref{workflow} we summarize the \texttt{RadFil} workflow, including a step-by-step overview of how \texttt{RadFil} builds and fits radial profiles. In \S \ref{architecture} we discuss the architecture of \texttt{RadFil}, describing how the package stores and summarizes the output. In \S \ref{relation} we provide a proof-of-concept, by reproducing filament width results derived from a completely independent, private radial density fitting code. Finally in \S \ref{future} we provide a roadmap for future developments, and how the functionality of \texttt{RadFil} can be extended and applied to diverse datasets.

\section{The \texttt{RadFil} Workflow} \label{workflow}
The \texttt{RadFil} workflow can be broken down into three stages: data input, filament profile building, and filament profile fitting. To illustrate the typical workflow for each stage, we have chosen a Herschel column density map for the Musca filament from \citet{Cox_2016} as an example. This filament is also used as a proof-of-concept in \S \ref{relation}. We discuss each stage in the \texttt{RadFil} workflow in more detail in subsections below.

\subsection{Data Input} \label{data_input}
\begin{figure}[h!]
\begin{center}
\includegraphics[width=1.0\columnwidth]{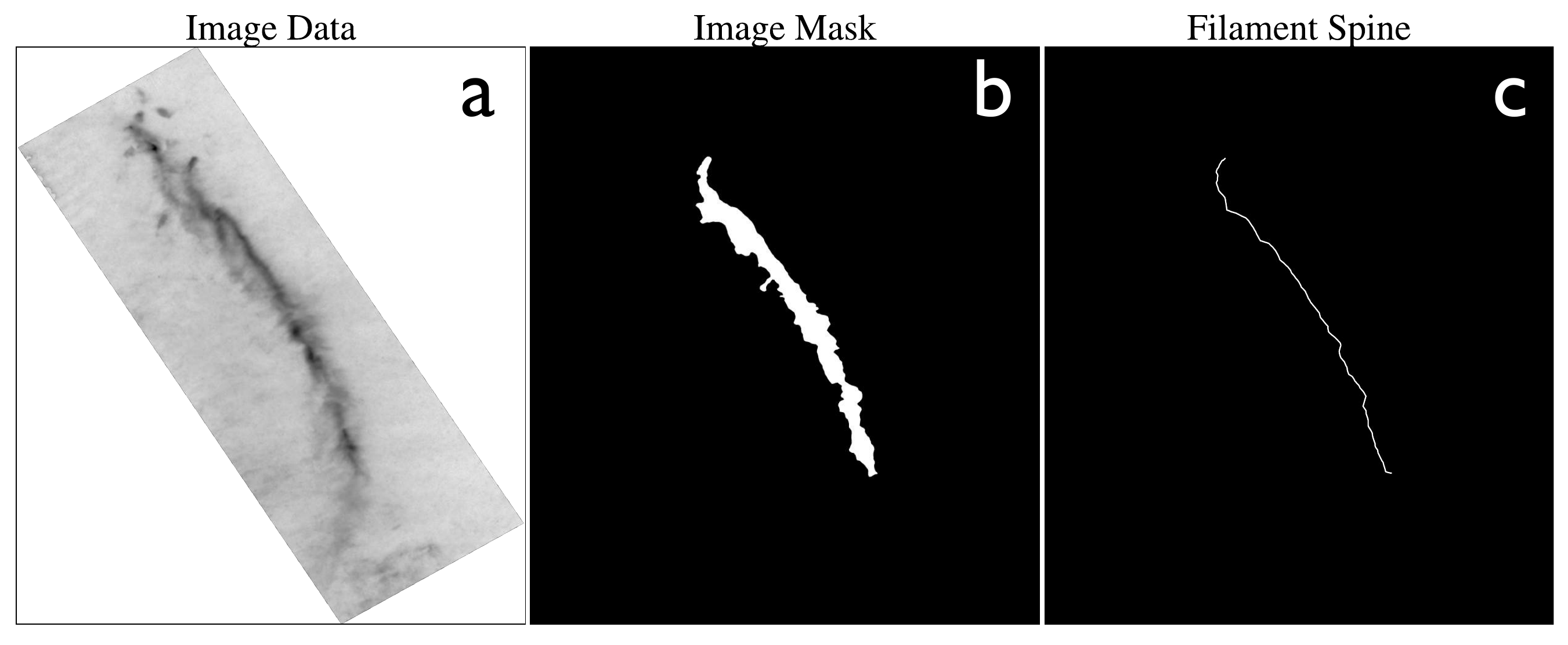}
\caption{\label{fig:musca_data} Step one in the typical \texttt{RadFil} workflow as summarized in \S \ref{data_input}. First, we input an image of the filament---in our example case, a map of the column density distribution of the Musca filament from the Herschel Gould Belt Survey \citep{Cox_2016}, as shown in Figure \textbf{(a)}. We additionally input a mask for the filament---see Figure \textbf{(b)}---calculated by applying a column density threshold $\approx 2 \sigma$ above the mean background column density of the image. This is equivalent to $\rm 2.25 \times 10^{21} \; cm^{-2}$. Finally, we create a filament spine---shown in \textbf{(c)}---by performing medial axis skeletonization on the filament mask, which produces a one-pixel one representation of the mask's topology. This skeletonization is performed using functionality from the \texttt{FilFinder} package \citep{Koch_2015}.}
\end{center}
\end{figure}

The first step in the \texttt{RadFil} workflow centers on data input and the code is designed to be flexible enough to accept filament data at any wavelength and at any resolution. All of the data are stored in a \texttt{radfil} class object and are loaded upon the instantiation of such an object. The only required datum is an image (in the form of a Python \texttt{numpy} array) which shows the intensity of the filament (see Figure \ref{fig:musca_data}a). This is typically read in as a FITS file. The corresponding FITS header for the image (containing for example, the WCS information) and the distance to the filament are optional inputs. If no header is provided the subsequent analysis will be carried out in pixel units. If a header is provided but not a distance, the analysis is carried out in angular units. If both are provided, the analysis will be carried out in physical units (i.e. parsecs). 

In addition to the image array, users can choose to input a filament mask (see Figure \ref{fig:musca_data}b) and/or a filament spine (both as \texttt{numpy} boolean arrays of the same shape as the image array), which are used in the profile building procedure described in \S \ref{profile_building}. The filament mask serves two purposes. First, in the case the user does not have a pre-built spine, the mask can be used to create a filament ``spine''---a one-pixel wide representation of the filament mask---using medial axis skeletonization\footnote{Also known as the medial axis transform, the medial axis skeletonization finds the set of all points within a shape that has more than one closest point on the shape's boundary.  The result is usually called the topological skeleton, or simply, the skeleton. In this case, for the input binary mask, we use \texttt{FilFinder} to find the skeleton(s) through the medial axis transform and derive the ``spine'' of the target filament based on the skeleton(s).} via the \texttt{FilFinder} package \citep{Koch_2015}. The medial axis transform produces skeletons by reducing the mask to a one-pixel wide binary representation of the mask's topology; the \texttt{FilFinder} algorithm then derives the shortest path from end-to-end using the set of skeletons to produce the final spine (see Figure \ref{fig:musca_data}c). This skeletonization procedure is performed via \texttt{RadFil}'s $make\_fil\_spine()$ command.\footnote{The $make\_fil\_spine()$ function is a wrapper function that calls the \texttt{FilFinder} package to perform medial axis skeletonization on the filament mask, and that package returns both the filament spine and its final length. There is no filament finding functionality in \texttt{RadFil} that is independent from \texttt{FilFinder}.} In addition to building the spines, in \S \ref{profile_building} the mask can also be used to define the extent of the image array for subsequent operations, including profile shifting (see \S\ref{profile_building}).

Alternatively, users have an option to input a pre-built spine via the ``filspine" argument \citep[for example, but not limited to, a spine from the \texttt{DisPerSE} program;][]{Sousbie_2013}. In this case, users can perform subsequent operations to obtain and fit the radial profile by running $build\_profile()$ and $fit\_profile()$, without running $make\_fil\_spine()$. See \S\ref{profile_building} and \S\ref{profile_fitting} for more details on obtaining and fitting the radial profile.

\subsection{Profile Building} \label{profile_building}
\begin{figure}[h!]
\begin{center}
\includegraphics[width=1.0\columnwidth]{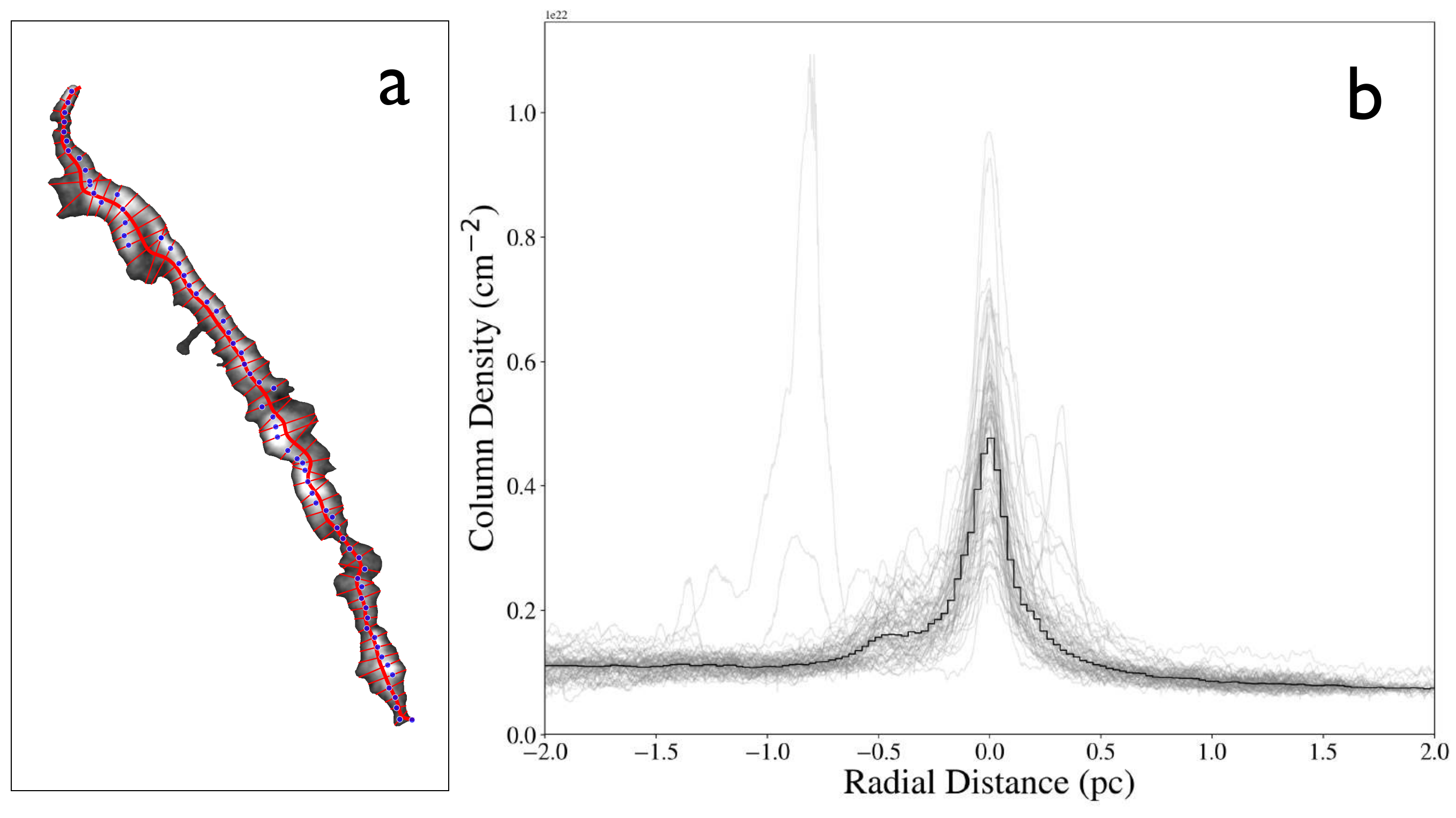}
\caption{\label{fig:musca_cuts} The profile building process for the Musca filament \citep{Cox_2016}, as described \S \ref{profile_building}. We take the filament spine shown in Figure \ref{fig:musca_data}c and smooth it via B-spline fitting. The smoothed spine is shown as the thick red line in Figure \textbf{(a)}, which has been overlaid on the column density distribution inside the filament mask (background grayscale). Then we take perpendicular cuts across the spine, which are shown as thin red intersecting lines in Figure (a). Finally, we shift each cut to the pixel with the peak column density, considering only those pixels along each cut and confined inside the filament mask. We mark the peak intensity pixels via blue scatter points in Figure (a). The column density profiles resulting from the cuts are shown as thin gray lines in Figure \textbf{(b)}. Users can also choose to bin the column density profiles of the individual cuts, and we show the binned average profile (with the ``bins" argument set to 600) with the black step function in Figure (b).}
\end{center}
\end{figure}

%% Fitting the B-spline
% Add pointers to the cartoon.
The second component of the \texttt{RadFil} workflow centers on building filament profiles using the $build\_profile()$ method. This is a multi-step process and the results of the profile building process for Musca \citep{Cox_2016} are summarized in Figure \ref{fig:musca_cuts}. First, we take the filament spine (a \texttt{numpy} mask of the same shape as the image array) which is either pre-computed or created as part of the data input process (\S\ref{data_input}). We then perform spline interpolation and fit a basis spline, ``B-spline",\footnote{A basis spline, or a ``B-spline,'' is a spline function with minimal support with respect to a given degree and smoothness.  Essentially, a B-spline represents the least complicated spline function---a function defined piecewise by polynomials---which satisfies the smoothness criterion in the interpolation.  The smoothness criterion is usually defined as $\sum_i{(w_i \, (S_i(x_i) - Y_i))^2} \leq s$, where $w_i$, $S_i$, and $Y_i$ are the weight, the spline function, and the input value at the position $x_i$, respectively, and $s$ is the smoothness. In reality, since the spine is not necessarily monotonic (i.e. there is not only one $y$ value for every $x$ value) we adopt a parametric representation and regress $x(t)$ and $y(t)$ along the spine, where $t$ represents the ordering of the pixels in the spine. To calculate the spine order, we first use a nearest neighbors algorithm to connect each ``node" (i.e. a pixel in the spine) to its two nearest neighbors, from which we construct a network graph. Based on the network graph, we construct a list of candidate paths, where each path goes through all nodes exactly one time. We then find the path with the lowest cost (i.e. distance) among all the candidate paths. Since paths starting at one end of the spine versus the other would in theory have the same cost, we parameterize the path such that it is always ordered by increasing $y$ value in the image} to the original spine points to determine a ``smoothed" continuous version of the spine. The positions and the first derivatives of the B-spline knots are used to create the cuts from which the profiles are built. Users can control how frequently cuts are made using the ``samp\_int'' argument---an integer which roughly specifies the sampling frequency in pixels. We emphasize this is an approximation. The knots which define the B-spline tend to be about one pixel length apart. This means that thinning the sampling points by the ``samp\_int'' parameter will produce an average physical distance between cuts of the sampling interval times the image scale of each pixel.  

\begin{figure}[h!]
\begin{center}
\includegraphics[width=0.7\columnwidth]{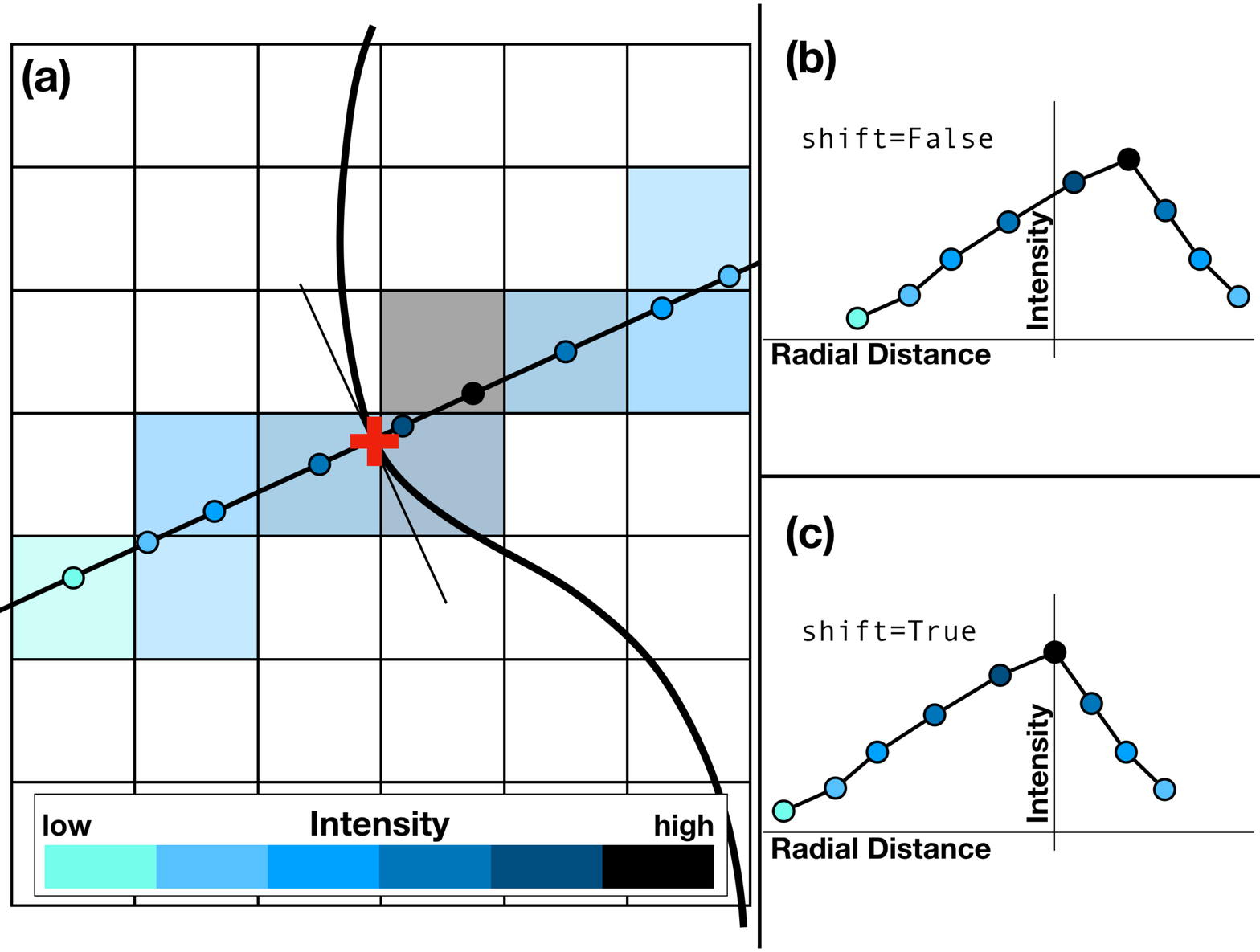}
\caption{\label{fig:cartoon} Schematic plots showing how the radial profile is calculated along a cut, and how profile shifting affects the profile. \textbf{(a)} The plot shows that the cut is derived based on the tangent at the sampling point (the red cross). \texttt{RadFil} calculates the radial distances from the spine and the intensities of the pixels that intersect it and stores the radial cut position that is closest to the middle of each pixel. \textbf{(b)} and \textbf{(c)} show the resulting profile along the cut shown in (a), when the shifting is set to False and True, respectively.  When the ``shift" argument is set to True, the radial profile is shifted such that the radial distance to the data point with the maximum intensity is 0.}
\end{center}
\end{figure}

%% Making cuts and derive the radial profiles.
At every sampling point (taken from the B-spline), \texttt{RadFil} parameterizes the line which is tangent to the curve at that point, using the first derivative. \texttt{RadFil} then finds the normal line---the line that is perpendicular to the tangent---and uses it to create the cut (Figure \ref{fig:cartoon}a). This process is repeated at every sampling point, creating an ensemble of cuts taken across the spine at approximately even intervals. Along each cut, \texttt{RadFil} calculates the radial distances from the spine and the intensities of the pixels that intersect it. Since the radial cut does not necessarily intersect neighboring pixels at their centers, we store the pixel positions corresponding to the midpoint of the cut's path through each pixel. We then use these positions to compute the radial distances (see Figure \ref{fig:cartoon}a). To determine the intensity at each radial distance, we perform bilinear grid interpolation using the set of samples along each cut. For Musca, the choice of interpolation scheme has a minimal effect on our results. If one adopts a ``nearest" interpolation scheme, as opposed to a bilinear one, it affects the best-fit values presented in \S4 at $<1\%$ level. 

%% Shift the profile.
\subsubsection{Optional: Profile Shifting}
While building the profiles, users also have the option to shift the radial profile along each cut so that it is centered on the pixel with the peak intensity. This is useful because the \texttt{FilFinder} skeletonization procedure often implemented to find the spine only takes into account the topology of the mask (and not the intensity distribution within it). Thus, the spine does not necessarily correspond to the peak intensity (e.g. in the case where there is an asymmetry in the radial profile with one side showing a substantial extended ``wing'' in its intensity distribution).

The option to shift the profile is controlled via the ``shift'' argument which accepts a boolean. If users choose to shift, \texttt{RadFil} will either search for the pixel with the peak intensity a) along each cut but inside the filament mask (if a mask is provided in \S \ref{data_input}) or b) along each cut but within a certain distance of the spine (if no mask is provided but a ``cutdist'' argument is given). Then it will recompute the radial distances and intensities assuming that pixel with the peak intensity (rather than the point of intersection with the spine) is situated at a radial distance of 0 pc (see Figures \ref{fig:cartoon}b and c). Because only points along the cut and confined inside the filament mask or within the ``cutdist" distance are considered, this ensures that the peak of each individual profile is centered on 0 pc while also ensuring that only physically meaningfully points are considered when attempting to determine where to shift the profile.

%% Bin the profile
\subsubsection{Optional: Profile Binning}
After the radial profiles along all the cuts are calculated, \texttt{RadFil} can optionally compute an averaged median profile for the filament by averaging the intensity values in radial distance bins whose number and width are controlled by the user. Users can specify a bin number via the ``bins'' argument which breaks up the profile into $n$ bins, spanning the minimum to maximum radial distances found in the ensemble of cuts. Alternatively, users can manually set the bin edges themselves, using, for example, the \texttt{numpy} linspace function, which creates an array of bin edges given some lower bound, upper bound, and number of bins. In more detail, \texttt{RadFil} sorts the ensemble of radial distance values into their appropriate radial distance bin, and then takes the median of their corresponding intensities to determine the average intensity at the center of the bin. If users choose to bin, fitting will be performed on the averaged binned profile in \S \ref{profile_fitting}. If no binning is applied, fitting can be performed on the entire ensemble of cuts without averaging. In Figure \ref{fig:musca_cuts}b, we show the entire ensemble of cuts without averaging via the thin gray lines, and overlay the averaged bin profile (with the ``bins" argument set to 600) as a black step function. 

%% Other options: pts_mask and fold.
\subsubsection{Optional: $pts\_mask$ and Profile Folding}
Finally, a few other $build\_profile()$ optional arguments are worthy of note. First, if there are any sections of the filament one would like excluded from the building process (so no samples are taken, etc.), one can enter the optional ``pts\_mask'' argument as a boolean mask of the same shape as the image array, which will discard any spine point that falls outside the region where ``pts\_mask=True.'' This can be useful in cases where the filament is experiencing stellar feedback (thereby disrupting its morphology), or when the filament loops back on itself (causing the same region to be sampled twice). Second, when taking the ``averaged'' profile via radial distance binning, users have the option to input a boolean value for the ``fold'' argument---this controls whether the averaged profile should be folded around the central radial distance, producing a half profile with only positive radial distance values. This is equivalent to averaging the intensities values within the same absolute radial distance bin from the spine (e.g. -3 and 3 pc) and assigning it to a distance of 3 pc. This is useful in cases where there is an asymmetry in the profile, which would affect the evaluation of the residual and could introduce biases to the fitting.

\subsection{Profile Fitting} \label{profile_fitting}
The final step in the \texttt{RadFil} workflow involves fitting the filament's ``master'' profile from the $profile\_building()$ step (see \S \ref{profile_building}). The ``master'' profile is dependent on the decisions users make in the $profile\_building()$ step regarding, for example, shifting, binning, and folding the profiles (see \S\ref{profile_building}).  All the fitting is performed via the $fit\_profile()$ function, which utilizes the \texttt{astropy.modeling} subpackage.

\subsubsection{Optional: Background Subtraction Prior to Profile Fitting}
 Prior to fitting the profile, \texttt{RadFil} supports background subtraction with a polynomial function of a non-negative degree. The degree of the polynomial is controlled via the ``bgdegree'' argument, and it accepts a non-negative integer including 0 (a zeroth order polynomial i.e. a constant) or 1 (a first order polynomial). The background fitting is done with the \texttt{astropy} implementation of a linear least-squares fitting routine.  In cases where the user would like to fit a higher-degree polynomial to the background, the linear least-squares fitting can be inefficient at finding the global minimum of the objective function.  Thus, users are expected to refrain from fitting a high-degree polynomial to the background, which could result in over-fitting and thus in loss of significant features after the fitted background is subtracted from the original radial profile.

Background subtraction is optional. However, if users choose to subtract a background, they can specify the range of radii over which to estimate the background. This is controlled via the ``bgdist'' argument, which accepts a tuple specifying the absolute lower and upper bounds of the background subtraction radii (e.g. bgdist=(1,2) indicates that the background will be estimated from -2 to -1 pc and from 1 to 2 pc). We perform basic outlier removal via sigma clipping, set to $3\sigma$. See Figure \ref{fig:plummer}a for an example of typical background subtraction. 

\subsubsection{Profile Fitting}
\begin{figure}[h!]
\begin{center}
\includegraphics[width=1.0\columnwidth]{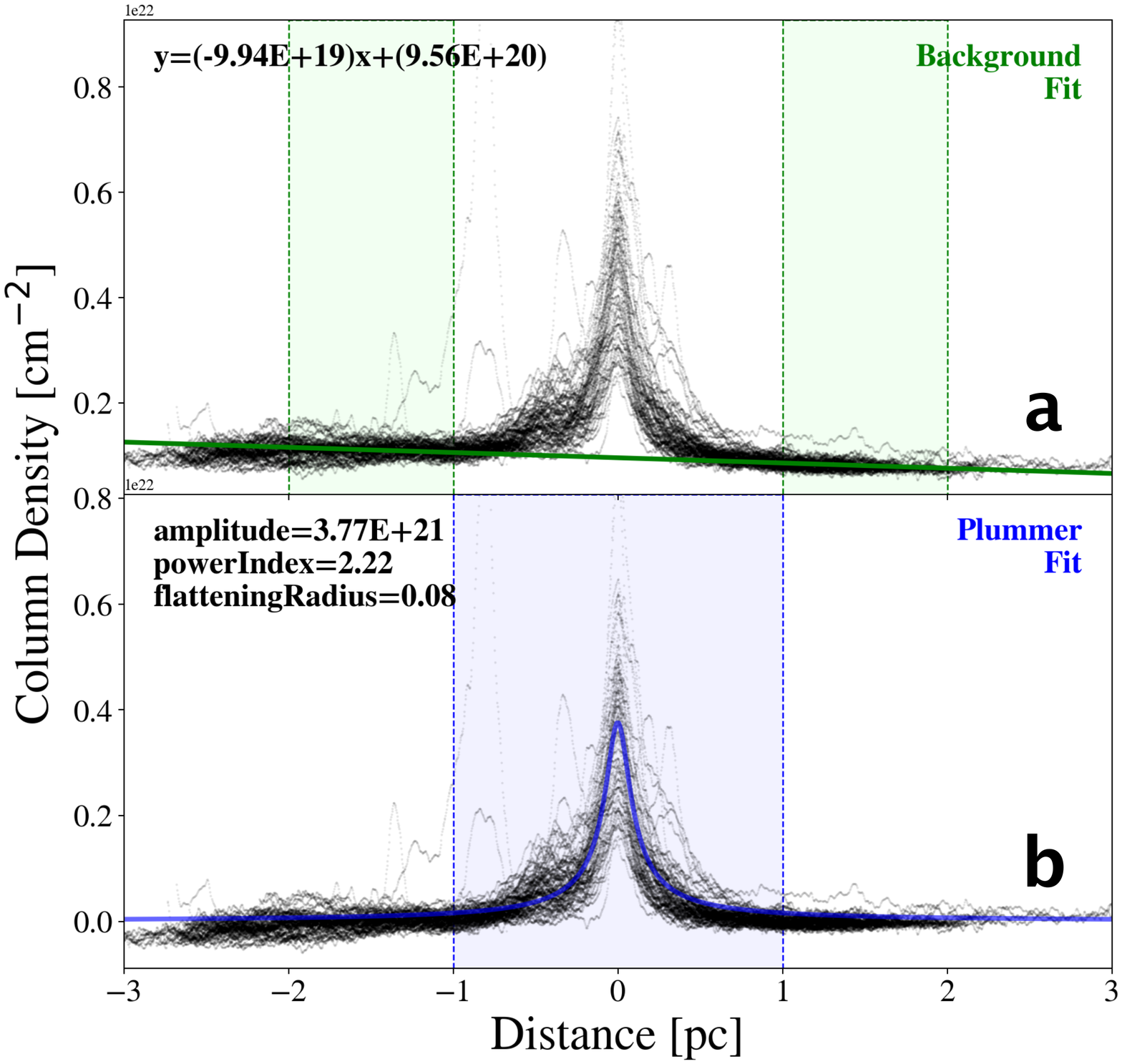}
\caption{\label{fig:plummer}  The results of applying our Plummer fitting procedure to the Musca filament from \citet{Cox_2016}, when adopting the same fitting radius (see also Figures \ref{fig:musca_data} and \ref{fig:musca_cuts}). In Figure \textbf{(a)}, we model the background as a first-order polynomial and estimate its contribution between 1 and 2 pc from the spine (green shaded regions, bounded by thin vertical green dashed lines). The best-fit background is shown as a green line in Figure (a) and the best-fit parameterization of that line is shown in the top left corner. This best-fit background is then subtracted off, producing the ensemble of radial profiles shown in Figure \textbf{(b)}. We fit a Plummer function to the subtracted profiles within 1.0 pc from the spine (blue shaded region, bounded by thin vertical blue dashed lines). The best-fit Gaussian profile is shown in Figure (b) in blue and the best-fit values are summarized in the top left corner.}
\end{center}
\end{figure}

\begin{figure}[h!]
\begin{center}
\includegraphics[width=1.0\columnwidth]{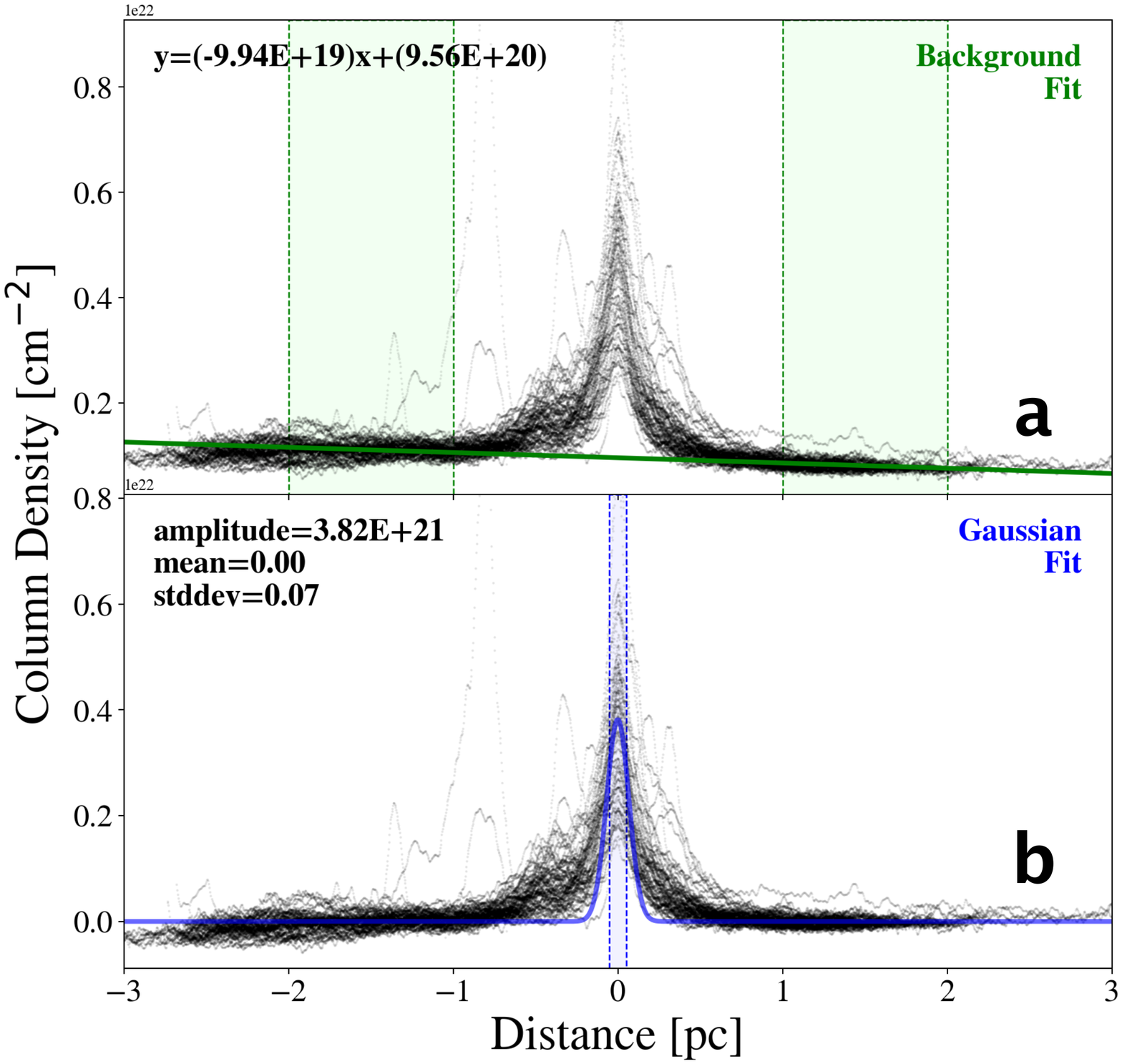}
\caption{\label{fig:gaussian} The results of applying our Gaussian fitting procedure to the Musca filament from \citet{Cox_2016}, when adopting the same fitting radius (see also Figures \ref{fig:musca_data} and \ref{fig:musca_cuts}). In Figure \textbf{(a)}, we model the background as a first-order polynomial and estimate its contribution between 1 and 2 pc from the spine (green shaded regions, bounded by thin vertical green dashed lines). The best-fit background is shown as a green line in Figure (a) and the best-fit parameterization of that line is shown in the top left corner. This best-fit background is then subtracted off, producing the ensemble of radial profiles shown in Figure \textbf{(b)}. We fit a Plummer function to the subtracted profiles within 0.05 pc from the spine (blue shaded region, bounded by thin vertical blue dashed lines). The best-fit Gaussian profile is shown in Figure (b) in blue and the best-fit values are summarized in the top left corner.}
\end{center}
\end{figure}

After choosing whether or not to background subtract, users have the option to fit two functions to the filament's profile---a Plummer (see Figure \ref{fig:plummer}a) or a Gaussian (see Figure \ref{fig:gaussian}a). The choice of function is specified via the ``fitfunc'' argument, which accepts a string (``Gaussian'' or ``Plummer''). Our Gaussian function is given by $H(r)=a \,\exp{(\frac{{-[r-\mu]}^2}{2\sigma^2})}$ where $r$ is the radial distance, $H$ is the profile height, and the free parameters are the amplitude ($a$), the standard deviation ($\sigma$) and optionally the mean ($\mu$). In the Gaussian case, users can decide whether to fit for the mean using the boolean argument ``fix\_mean," which fixes the mean to zero when True and lets it float in the fit when False. The default behavior is to set ``fix\_mean=True" when ``shift=True" and ``fix\_mean=False" otherwise. Our Plummer function is the same as that found in \citet{Cox_2016} and is given by $H(r) = \frac{H_0}{[{1+({\frac{r}{R_{flat}})^2}]}^{\; \frac{p-1}{2}}}$, where $H_0$ is the peak profile height, $R_{flat}$ is the flattening radius, and $p$ is the index of the density profile. When fitting the Plummer, we assume that the inclination angle of the filament to the plane-of-the-sky is zero. 

\texttt{RadFil} uses the \texttt{astropy} implementation of the Levenberg-Marquardt algorithm to fit the profile.  The Levenberg-Marquardt algorithm can be seen as an interpolation between the Gauss-Newton algorithm and the gradient descent, and it is preferable in non-linear curve fitting, especially in cases where the initial guess cannot be determined without a large uncertainty.  As in the background fitting, the objective function is the least-squares residual. \texttt{RadFil} currently only supports fitting to the ``master'' profile (e.g. the entire ensemble of cuts in the case of no binning, the binned median profile in the case of binning, or the ``folded'' version of either of these two cases). However, users can easily export the radial distances and intensities for each cut individually, to which users can apply their own fitting routines (see \S \ref{architecture} for more information on how \texttt{RadFil} organizes its output).

Just like when subtracting the background, \texttt{RadFil} allows the user to specify the fitting radii over which to perform the fit. This is critical, as several studies have shown that the best-fit parameters for Gaussian and Plummer functions are extremely sensitive to choice of both fitting distance and background subtraction radii \citep{Smith_2014, Zucker_2017, Panopoulou_2017}. The fitting distance is controlled via the ``fitdist'' argument, which accepts either a float or a tuple. If a float is entered, \texttt{RadFil} will fit to that radial distance symmetrically on either side of the spine. However, if users would like an asymmetric fitting range, they can enter a tuple with a lower and upper bound (e.g. -1 to 2 pc) and \texttt{RadFil} will only consider that range of radial distances when performing the fit. 

Finally, \texttt{RadFil} is capable of computing the deconvolved widths if a beamwidth for the image is provided. If possible, \texttt{RadFil} will automatically read the beamwidth keyword from the FITS header. Otherwise, users can provide a value in arcseconds for the ``beamwidth'' argument manually in $fit\_profile()$. If no beamwidth is provided, \texttt{RadFil} will only compute the convolved FWHM value for the filament, using the formula $FWHM=2 \sqrt{2\log(2)} \, \sigma$, where $\sigma$ is the standard deviation when a Gaussian function is chosen and the flattening radius when a Plummer function is chosen. If a beamwidth is provided, \texttt{RadFil} determines the deconvolved FWHM using the following formula $FWHM_{deconv}=\sqrt{FWHM^2-B^2}$ where B is the physical beamwidth and FWHM is the convolved FWHM computed above. This definition of the deconvolved FWHM is taken from \citet{Konyves_2015}. \textit{This deconvolution procedure is \textit{not} robust}. Convolving the data with a beam of known size and recalculating the deconvolved FWHM produces different values. To illustrate this, we have taken the original Musca column density map (see Figure \ref{fig:musca_data}a), with a beamwidth of $36.3\arcsec$ (0.035 pc), and degraded the resolution of the map by a factor of two and four, so the new maps have a beamwidth of $1.2\arcmin$ (0.070 pc) and $2.4\arcmin$ (0.14 pc), respectively. Adopting the same fitting parameters as before, we determine the deconvolved FWHM values in both cases, the results of which are reported in Table \ref{tab:FWHM}. For the Gaussian fits, the deconvolved FWHM for the $2.4\arcmin$ map is $1.6\times$ higher than for the $36\arcsec$ map. The Plummer fits appear slightly more robust, producing a deconvolved FWHM that is only $1.3\times$ higher. Users should be aware that the deconvolved FWHM can increase significantly as the resolution of the map decreases. Nevertheless, this is the deconvolution procedure adopted throughout the literature \citep[see, for instance,][]{Panopoulou_2017, Koch_2015}.  

\subsubsection{Uncertainties in Best-Fit Parameter Values} \label{uncertainties}
There are two types of uncertainties inherent in \texttt{RadFil's} fitting procedure. First, there is statistical uncertainty intrinsic to our least-squares fitting algorithm. Our choice of fitter, \texttt{astropy's} \href{http://docs.astropy.org/en/stable/api/astropy.modeling.fitting.LevMarLSQFitter.html}{LevMarLSQFitter}, returns a covariance matrix, and the statistical uncertainties on the best-fit values can be calculated by taking the square-root of the diagonal elements of this matrix \citep{Hogg_2010}. After running the \textit{fit\_profile()} method, users can access these uncertainties via \texttt{RadFil's} ``std\_error" attribute, which returns an array whose length is equal to the number of fitted parameters. An alternative method is to calculate the statistical uncertainty empirically, using either bootstrap or jackknife resampling methods \citep[see discussion in Section 4 of][]{Hogg_2010}. We plan to add support for empirical error calculations in the future. 

In additional to statistical uncertainty, there is also systematic uncertainty stemming from the user's choice of fitting parameters. For reasonably built filament profiles (whose cuts are well-sampled but also statistically independent) the choice of background subtraction and fitting radii tend to dominate the systematic uncertainty. The ``bgdegree" argument (controlling whether a zeroeth or first order polynomial is fit to the background) and the ``fix\_mean" argument (controlling whether the mean is fixed to zero in the Gaussian case) only become important if the data are highly skewed. To quantify the uncertainty based on background subtraction and fitting radii, \texttt{RadFil} has a built-in function called \textit{calculate\_systematic\_uncertainty()} which accepts two lists, one containing various options for the background subtraction radii, and the other containing options for the fitting radius. \texttt{RadFil} will compute all possible combinations of background subtraction and fitting radii given these two lists, and then determine the best-fit values in each case for either a Gaussian or Plummer model. RadFil will then store a dictionary (``radfil\_trials"), with each dictionary key corresponding to a different parameter in the model. Accessing these keys will return a \texttt{pandas} dataframe (where the columns are the fitting radii and the rows are the background subtraction radii), which users can easily print to screen to summarize the results.

\begin{table}[ht!]
\resizebox{\linewidth}{!}{% Resize table to fit within \linewidth horizontally
\hskip-2cm \begin{tabular}{|c|c|c|c|c|}
\hline
\multirow{2}{*}{Beamwidth} & \multicolumn{2}{c|}{Gaussian} & \multicolumn{2}{c|}{Plummer} \\ \cline{2-5} 
                           & FWHM (pc)  & $\rm FWHM_{deconv}$ (pc)  & FWHM (pc)  & $\rm FWHM_{deconv}$ (pc) \\ \hline
36.3\arcsec                      & 0.159      & 0.155            & 0.221      & 0.218           \\ \hline
1.2\arcmin                       & 0.197      & 0.184            & 0.248      & 0.238           \\ \hline
2.4\arcmin                       & 0.282      & 0.244            & 0.321      & 0.289           \\ \hline
\end{tabular}%
}
\caption{{ Dependence of deconvolved FWHM on the beamwidth of the image data. \texttt{RadFil} defines the $FWHM_{deconv}=\sqrt{FWHM^2-B^2}$ \citep{Konyves_2015} where B is the physical beamwidth of the image and $FWHM=2 \sqrt{2\log(2)} \, \sigma$. The $\sigma$ parameter corresponds to the standard deviation when a Gaussian function is chosen and the flattening radius when a Plummer function is chosen. Adopting a distance to Musca of 200 pc \citep{Cox_2016}, a beamwidth of $36\arcsec$, $1.2\arcmin$, and $2.4\arcmin$ corresponds to a physical size of 0.035 pc, 0.070 pc, and 0.14 pc, respectively. The deconvolution procedure \texttt{RadFil} adopts from \citet{Konyves_2015} is \textit{not} robust---the deconvolved FWHM increases as the resolution of the image decreases.
{\label{tab:FWHM}}}}
\end{table}

\section{The \texttt{RadFil} Architecture} \label{architecture}
When users input data (\S \ref{data_input}), build their filament profiles (\S \ref{profile_building}), and fit their filament profiles (\S \ref{profile_fitting}), \texttt{RadFil} stores all the relevant information computed at each step as attributes of the \texttt{radfil} class object, which users can access at any time or export for further processing. 

For instance, once a \texttt{radfil} class object is instantiated, any data or properties that the user inputs into \texttt{RadFil} (e.g. the filament image, mask, spine, distance, etc.) are accessible by calling the name of the \texttt{radfil} class object (e.g. ``radobj'') and then the name of the input argument. For instance, users can type radobj.image, radobj.distance, or radobj.filspine to return any of these arrays or quantities. If users choose to make their own filament spine using \texttt{FilFinder} and \texttt{RadFil's} $make\_fil\_spine()$ function, users can also return the length of that spine by calling radobj.length.  

Once users have built their profiles, a new set of attributes becomes available. For instance, users can access the ordered list of the x and y pixel coordinates defining the spine points (both before and after the smoothing via the B-spline fitting) by calling radobj.xbeforespline and radobj.ybeforespline, or radobj.xspline and radobj.yspline respectively. Most importantly, \texttt{RadFil} also stores all the radial distances and profile heights of individual cuts across the spine as a dictionary, which users can access by calling the ``dictionary\_cuts'' attribute. The dictionary includes keys like ``profile'' and ``distance'', both of which access a list containing a set of \texttt{numpy} arrays, where the number of arrays is equivalent to the number of cuts across the spine. Each individual array contains the radial distances or the profile heights corresponding to each cut. Another dictionary key of interest is the ``mask\_width'' key which contains a list of floats, where each float indicates the width of the individual cut confined within the mask (i.e. the distance between where the cut intersects the edge of the mask on either side of the spine). The average of these values can then be used to estimate a ``mask-based'' width of the filament (e.g. in the case of Musca, the average width of some closed contour). This is equivalent to taking the average of all the lengths of the thin red cuts intersecting the filament spine in Figure \ref{fig:musca_cuts}a. 

Finally, once users have performed background subtraction and/or fit their profile, they can access the resulting best-fit parameters via the radobj.bgfit and radobj.profilefit attributes, which will return the \texttt{astropy.modeling} objects. These modeling objects will then have their own attributes users can access to get information about the fit (i.e. the attributes ``param\_names" or ``parameters" return the variable names and the best-fit parameter values, respectively). As one avenue to calculate the uncertainty on the best-fit parameters, users can access the covariance matrix for the fit using the command radobj.param\_cov. This covariance matrix is used to calculate the standard errors on the best-fit parameters, which can be accessed via radobj.std\_error. However, we strongly caution that using the covariance matrix will likely underestimate the errors, as simply tweaking knobs (like the fitting radius) will produce uncertainties in the best-fit parameters much larger than those derived from the covariance matrix. To quantify the systematic uncertainties, users can run \textit{calculate\_systematic\_uncertainty()} (as discussed in \S \ref{uncertainties}), which will return a set of \texttt{pandas} dataframes summarizing the various best-fit values given different background subtraction and fitting radii. Finally, users can obtain the FWHM of the filament and (if a beamwidth is entered) the deconvolved FWHM using the radobj.FWHM and radobj.FWHM\_deconv commands.

\section{Proof-of-Concept} \label{relation}
As a proof-of-concept, we have tested the results \texttt{RadFil} provides against the same results produced by a private and independent radial density profile fitting code. For a fair comparison, we apply our technique to an identical dataset and adopt the same fitting parameters. To reiterate, for this purpose we have chosen the Musca filament as presented in \citet{Cox_2016}, whose column density map is publicly available online as part of the \href{http://www.herschel.fr/cea/gouldbelt/en/Phocea/Vie_des_labos/Ast/ast_visu.php?id_ast=66}{Herschel Gould Belt Survey archive}. This filament is also used as an example in the \texttt{RadFil} tutorial, which is likewise available \href{https://github.com/catherinezucker/radfil/blob/master/RadFil_Tutorial.ipynb}{online as a Jupyter notebook} that users can download and run themselves. We adopt the same 200 pc distance to Musca as in \citet{Cox_2016}. 

We first start by applying a column density threshold to the publicly available Musca column density map ($\approx 2\sigma$ above the background column density or $\rm 2.25 \times 10^{21} \; cm^{-2}$), in order to produce a mask which delineates the morphology of the filament. The column density map is shown in Figure \ref{fig:musca_data}a while the resulting filament mask is shown in Figure \ref{fig:musca_data}b. We then run the $make\_fil\_spine()$ command to produce a filament spine via medial axis skeletonization, as discussed in \citet{Koch_2015}. This spine is shown in Figure \ref{fig:musca_data}c. Using the \texttt{FilFinder} package we obtain a length of 8.1 pc, which is in excellent agreement with the length found in \citet{Cox_2016} ($\approx 8$ pc, see their Figure 4a). 

When building the profile, we adopt a pixel sampling interval (``samp\_int'') of 25, equivalent to making a cut approximately every 0.10 pc given the image scale of the Musca map ($\rm \frac{0.00387 \; pc}{pixel}$). In Figure \ref{fig:musca_cuts}a, these cuts are shown as thin red lines intersecting the smoothed filament spine (obtained via B-spline fitting), which is plotted in thick red. The column density distribution inside the filament mask is shown in grayscale in the background. After making the cuts, we choose to shift the filament profile to the peak intensity along each cut confined inside the filament mask. The point of peak intensity across each cut is marked via a blue scatter point in Figure \ref{fig:musca_cuts}a. Each individual cut is shifted so that this position is assigned a radial distance of zero, and the profile is now built out from this point. The individual shifted profiles corresponding to the cuts are shown in gray in Figure \ref{fig:musca_cuts}b. 

As in \citet{Cox_2016}, we fit both a Gaussian and a Plummer function. We first subtract a first-order linear background, estimated between 1 and 2 pc on either side of the spine (Figures \ref{fig:plummer}a and \ref{fig:gaussian}a). For their Plummer function \citet{Cox_2016} choose a fitting distance of 1 pc (N. Cox; private communication), determining a best-fit flattening radius of 0.08 pc and power index of $2.2\pm 0.3$ \citep[see Figure 4b in][]{Cox_2016}. Fitting to the ensemble of cuts (Figure \ref{fig:musca_cuts}b) out to an identical fitting radius, we determine a best-fit flattening radius of 0.08 pc and best-fit power index of 2.22. Our best-fit Plummer function is shown in Figure \ref{fig:plummer}b. The statistical uncertainty on these parameters, computed from the covariance matrix, is low, equivalent to 0.001 pc for the flattening radius and 0.01 for the power index. To calculate the systematic uncertainty, we use \texttt{RadFil's} \textit{calculate\_systematic\_uncertainty()} function. We test fitting radii of 0.5 pc, 1.0 pc, 1.5 pc, and 2.0 pc, and background subtraction radii of 1-2, 2-3 pc, 2-4 pc, and 3-4 pc. The results are summarized in Table \ref{tab:plummer_systematics}. By taking the standard deviation of the set of best-fit values for each parameter (Table \ref{tab:plummer_systematics}), we find that the systematic uncertainty is about an order of magnitude higher than the statistical uncertainty, or 0.01 pc for the flattening radius and 0.13 for the power index. Adding the uncertainties in quadrature, we determine $R_{flat}= 0.08 \pm 0.01$ pc and $p=2.22\pm 0.13$, in strong agreement with the \citet{Cox_2016} results. 

Next, for their Gaussian function \citet{Cox_2016} choose a fitting distance of 0.05 pc (N. Cox; private communication) and obtain a deconvolved FWHM width of $0.14 \pm 0.03$ pc. Again, adopting this same cutoff, we fit a Gaussian to our ensemble of cuts (Figure \ref{fig:musca_cuts}b) and obtain a best-fit deconvolved FWHM value of 0.16 pc, with a statistical uncertainty of 0.008 pc. Our best-fit Gaussian function is shown in Figure \ref{fig:gaussian}b. However, similar to the Plummer fits, the systematic uncertainties dominate. To calculate the systematic uncertainty, we again use \texttt{RadFil's} \textit{calculate\_systematic\_uncertainty()} function. We test fitting radii of 0.05 pc, 0.1 pc, 0.25 pc, 0.5 pc, and 1 pc, and background subtraction radii of 1-2 pc, 2-3 pc, 2-4 pc, and 3-4 pc. The results are summarized in Table \ref{tab:gaussian_systematics}. By taking the standard deviation of the set of best-fit values for the width, we calculate a systematic uncertainty on the FWHM value of 0.09 pc. Adding the uncertainties in quadrature, we calculate a deconvolved FWHM of $0.16 \pm 0.09$ pc, which is again consistent with the published value from \citet{Cox_2016}. While the systematic uncertainties we calculate depend on our background subtraction radii and fitting radii input lists, it does highlight how a wide range of best-fit values are attainable simply by tweaking the fitting knobs. Thus, it is important to be explicit about which fitting parameters one adopts when reporting best-fit values obtained from radial profile fitting.

% \usepackage{multirow}
\begin{table}[]
\resizebox{18cm}{!}{
\hskip-7cm 
\renewcommand{\arraystretch}{2.5}
\begin{tabular}{|c|c|c|c|c|c|c|c|c|c|c|c|c|c|c|c|c|c|}
\hline
\multicolumn{2}{|c|}{\multirow{3}{*}{\shortstack[c]{\huge Amplitude \\ \large ($\rm cm^{-2})$}}} & \multicolumn{4}{c|}{\multirow{2}{*}{\large Fitting Radius}} & \multicolumn{2}{c|}{\multirow{3}{*}{\shortstack[c]{\huge Flattening \\ \huge Radius \\ \large (pc)}}} & \multicolumn{4}{c|}{\multirow{2}{*}{\large Fitting Radius}} & \multicolumn{2}{c|}{\multirow{3}{*}{\shortstack[c]{\huge Power \\ \huge Index}}} & \multicolumn{4}{c|}{\multirow{2}{*}{\large Fitting Radius}} \\
\multicolumn{2}{|c|}{} & \multicolumn{4}{c|}{} & \multicolumn{2}{c|}{} & \multicolumn{4}{c|}{} & \multicolumn{2}{c|}{} & \multicolumn{4}{c|}{} \\ \cline{3-6} \cline{9-12} \cline{15-18} 
\multicolumn{2}{|c|}{} & 0.5 pc & 1 pc & 1.5 pc & 2 pc & \multicolumn{2}{c|}{} & 0.5 pc & 1 pc & 1.5 pc & 2 pc & \multicolumn{2}{c|}{} & 0.5 pc & 1 pc & 1.5 pc & 2 pc \\ \hline
\multirow{4}{*}{\shortstack[c]{\large Background \\ \large Subtraction \\ \large Radii}} & 1-2 pc & 3.84E+21 & 3.77E+21 & 3.72E+21 & 3.70E+21 & \multirow{4}{*}{\shortstack[c]{\large Background \\ \large Subtraction \\ \large Radii}} & 1-2 pc & 0.063 & 0.076 & 0.084 & 0.089 & \multirow{4}{*}{\shortstack[c]{\large Background \\ \large Subtraction \\ \large Radii}} & 1-2 pc & 2.05 & 2.22 & 2.34 & 2.3 \\ \cline{2-6} \cline{8-12} \cline{14-18} 
 & 2-3 pc & 3.86E+21 & 3.79E+21 & 3.75E+21 & 3.74E+21 &  & 2-3 pc & 0.063 & 0.074 & 0.082 & 0.084 &  & 2-3 pc & 2.03 & 2.19 & 2.28 & 2.31 \\ \cline{2-6} \cline{8-12} \cline{14-18} 
 & 2-4 pc & 3.83E+21 & 3.76E+21 & 3.71E+21 & 3.70E+21 &  & 2-4 pc & 0.064 & 0.077 & 0.085 & 0.087 &  & 2-4 pc & 2.05 & 2.23 & 2.34 & 2.37 \\ \cline{2-6} \cline{8-12} \cline{14-18} 
 & 3-4 pc & 3.82E+21 & 3.74E+21 & 3.69E+21 & 3.68E+21 &  & 3-4 pc & 0.064 & 0.078 & 0.087 & 0.089 &  & 3-4 pc & 2.07 & 2.26 & 2.39 & 2.41 \\ \hline
\end{tabular}
}
\caption{\label{tab:plummer_systematics} The effect of tweaking the background subtraction radii and fitting radius on our Plummer model parameters, with amplitude (left), flattening radius (middle), and index of the density profile (right). To estimate the background, we fit a first-order polynomial between radial distances of 1-2 pc, 2-3 pc, 2-4 pc, and 3-4 pc. We adopt fitting radii of 0.5 pc, 1 pc, 1.5 pc, and 2 pc. For each combination we compute the best-fit amplitude, flattening radius, and power index to quantify the systematic effect that tuning these parameters has on our results. This is performed via \texttt{RadFil's} \textit{calculate\_systematic\_uncertainty()} function.}
\end{table}

\begin{table}[]
\resizebox{18cm}{!}{
\hskip-5cm 
\renewcommand{\arraystretch}{2}
\begin{tabular}{|c|c|c|c|c|c|c|c|c|c|c|c|c|c|}
\hline
\multicolumn{2}{|c|}{\multirow{3}{*}{\shortstack[c]{\huge Amplitude \\ \large ($\rm cm^{-2})$}}} & \multicolumn{5}{c|}{\multirow{2}{*}{\large Fitting Radius}} & \multicolumn{2}{c|}{\multirow{3}{*}{\shortstack[c]{\huge Standard\\ \huge Deviation \\ \large (pc)}}} & \multicolumn{5}{c|}{\multirow{2}{*}{\large Fitting Radius}} \\
\multicolumn{2}{|c|}{} & \multicolumn{5}{c|}{} & \multicolumn{2}{c|}{} & \multicolumn{5}{c|}{} \\ \cline{3-7} \cline{10-14} 
\multicolumn{2}{|c|}{} & 0.05 pc & 0.1 pc & 0.25 pc & 0.5 pc & 1 pc & \multicolumn{2}{c|}{} & 0.05 pc & 0.1 pc & 0.25 pc & 0.5 pc & 1 pc \\ \hline
\multirow{4}{*}{\shortstack[c]{\large Background \\ \large Subtraction \\ \large Radii}}& 1-2 pc & 3.82E+21 & 3.70E+21 & 3.29E+21 & 2.96E+21 & 2.94E+21 & \multirow{4}{*}{\shortstack[c]{\large Background \\ \large Subtraction \\ \large Radii}} & 1-2 pc & 0.068 & 0.083 & 0.12 & 0.16 & 0.16 \\ \cline{2-7} \cline{9-14} 
 & 2-3 pc & 3.84E+21 & 3.72E+21 & 3.31E+21 & 2.96E+21 & 2.93E+21 &  & 2-3 pc & 0.068 & 0.084 & 0.12 & 0.16 & 0.16 \\ \cline{2-7} \cline{9-14} 
 & 2-4 pc & 3.82E+21 & 3.70E+21 & 3.29E+21 & 2.96E+21 & 2.94E+21 &  & 2-4 pc & 0.068 & 0.083 & 0.12 & 0.16 & 0.16 \\ \cline{2-7} \cline{9-14} 
 & 3-4 pc & 3.81E+21 & 3.68E+21 & 3.28E+21 & 2.96E+21 & 2.95E+21 &  & 3-4 pc & 0.067 & 0.083 & 0.12 & 0.15 & 0.15 \\ \hline
\end{tabular}
}
\caption{\label{tab:gaussian_systematics} The effect of tweaking the background subtraction radii and fitting radius on our Gaussian model parameters, with amplitude (left) and standard deviation (right). To estimate the background, we fit a first-order polynomial between radial distances of 1-2 pc, 2-3 pc, 2-4 pc, and 3-4 pc. We adopt fitting radii of 0.05 pc, 0.1 pc, 0.25 pc, 0.5 pc, and 1.0 pc. For each combination we compute the best-fit amplitude and standard deviation to quantify the systematic effect that tuning these parameters has on our results. This is performed via \texttt{RadFil's} \textit{calculate\_systematic\_uncertainty()} function.}
\end{table}

\section{A Roadmap for Future Development} \label{future}
\texttt{RadFil} is undergoing continuous development and all the code is currently publicly available at \href{https://github.com/catherinezucker/radfil}{this GitHub repository}. The code used to produce the results in this paper has also been archived on Zenodo with a unique DOI (\url{https://doi.org/10.5281/zenodo.1287318}). The package is also available on \href{https://pypi.python.org/pypi} {PyPi} and can be installed with pip. Since \texttt{RadFil} is open-source, users are welcome to clone the repository on GitHub and make any changes to the code that they desire. Likewise they are encouraged to report any issues or feature requests via the issue tracker on GitHub. The code itself can be run in Python 2.7, 3.4, 3.5, or 3.6. 

At the moment \texttt{RadFil} is only designed to work with 2D images. Current datasets ripe for analysis include maps of dust emission \citep[using e.g. the Hi-GAL survey;][]{Molinari_2016} and dust extinction \citep[using e.g. the GLIMPSE survey;][]{Churchwell_2009}, or else integrated intensity maps from a wide-variety of spectral-lines \citep[for example $\rm ^{13}CO, C^{18}O$ from surveys like GRS, SEDIGISM, or COHRS;][]{Jackson_2006, Schuller_2017, Dempsey_2013}. Nor is \texttt{RadFil} limited to observational data. As long as the data are on a regular grid, one could apply \texttt{RadFil} to 2D slices obtained from 3D numerical simulations, or integrated intensity maps from synthetic spectral cubes.

In the coming years we hope to expand \texttt{RadFil's} functionality so that it can be applied to 3D datasets, most notably \textit{position-position-velocity (p-p-v)} spectral cubes from both low and high density gas tracers, or else 3D density cubes obtained from hydrodynamical simulations. Existing structure identification packages like $DisPerSE$ \citep{Sousbie_2013} already have the capability to extract filaments in 3D and have been applied to cosmic web filaments in dark matter simulations \citep{Sousbie_2011} as well as star-forming filaments in gravoturbulent molecular cloud simulations \citep{Klassen_2017}.
While still under development, the \texttt{FilFinder} algorithm \citep{Koch_2015} also has the ability to extend its filament identification procedure to three dimensions, and \citet{Koch_2015} cite the future application of their algorithm to 3D \textit{p-p-v} spectral cubes as a primary motivation. We envision that \texttt{RadFil} would be a natural extension of these filament identification packages, by providing additional functionality and flexibility for users interested in more robust width determination. In this vein, we hope to continue to improve the ease by which \texttt{RadFil} can be integrated into existing filamentary analysis pipelines, and we welcome community investment in its development both now and in the future. 

\acknowledgments
We would like to thank Cara Battersby and Alyssa Goodman for their valuable expertise that helped to bring \texttt{RadFil} to fruition. The original idea for \texttt{RadFil} came from Cara Battersby and was inspired by the analysis she presented in \citet{Battersby_2014}. Likewise the original idea for the publication of this software paper came from Alyssa Goodman.

\software{Astropy \citep{astropy_collaboration}, Matplotlib \citep{matplotlib}, FilFinder \citep{filfinder}, Scikit-learn \citep{scikit-learn}, Scikit-image \citep{scikit-image}, Networkx \citep{networkx}, Pandas \citep{pandas}}

\bibliography{main_revision_v2.bib}

\end{document}